\documentclass[aps, twocolumn,superscriptaddress]{revtex4-1}
\usepackage{graphicx} 
\usepackage{amssymb,amsfonts,amsmath}
\usepackage{txfonts}
\usepackage[font=small]{caption} 
\usepackage{xcolor}              
\usepackage{setspace}            
\usepackage{fullpage}            
\usepackage{url}
\usepackage[justification=justified,singlelinecheck=false]{subcaption}
\usepackage{booktabs}
\usepackage{url}
\usepackage{placeins}
\usepackage{epstopdf}
\usepackage{verbatim}

\definecolor{darkgreen}{HTML}{00BB00}

\begin{document}
\title{Resilience of Networks Formed of Interdependent Modular Networks}
\date{\today}
\author{Louis M. Shekhtman}
\affiliation{Department of Physics, Bar-Ilan University, Ramat Gan, Israel}
\author{Saray Shai}
\affiliation{Department of Mathematics, University of North Carolina, Chapel Hill, North Carolina, USA}
\author{Shlomo Havlin}
\affiliation{Department of Physics, Bar-Ilan University, Ramat Gan, Israel}

\begin{abstract} 
	Many infrastructure networks have a modular structure and are also interdependent with other infrastructures. While significant research has explored the resilience of interdependent networks, there has been no analysis of the effects of modularity.
	Here we develop a theoretical framework for attacks on interdependent modular networks and support our results through simulations.
	We focus, for simplicity, on the case where each network has the same number of communities and the dependency links are restricted to be between pairs of communities of different networks. This is particularly realistic for modeling infrastructure across cities. Each city has its own infrastructures and different infrastructures are dependent only within the city. However, each infrastructure is connected within and between cities. For example, a power grid will connect many cities as will a communication network, yet a power station and communication tower that are interdependent will likely be in the same city. 
	It has previously been shown that single networks are very susceptible to the failure of the interconnected nodes (between communities) \cite{shai2014resilience} and that attacks on these nodes are even more crippling than attacks based on betweenness \cite{da2015complex}. In our example of cities these nodes have long range links which are more likely to fail. 
	For both treelike and looplike interdependent modular networks we find distinct regimes depending on the number of modules, $m$. (i) In the case where there are fewer modules with strong intraconnections, the system first separates into modules in an abrupt first-order transition and then each module undergoes a second percolation transition. (ii) When there are more modules with many interconnections between them, the system undergoes a single transition.
	Overall, we find that modular structure can significantly influence the type of transitions observed in interdependent networks and should be considered in attempts to make interdependent networks more resilient.
\end{abstract}
\maketitle

\section{Introduction}
Many recent studies have explored interdependent  and multilayer networks  \cite{ buldyrev-nature2010, gao-naturephysics2012, parshani-prl2010,  leichtdsouza2009, cellai-pre2013, brummitt-pnas2012,  PhysRevE.90.012803, gao-prl2011, hu-pre2011, bashan-pre2011, parshani-pnas2011, bashan-jsp2011, PhysRevE.85.066134, vespignani-nature2010,  rinaldi-ieee2001, peerenboom-proceedings2001, radicchi-naturephysics2013, son-epl2012, zhao-jstatmech2013, donges-epjb2011, baxter-prl2012, gomez-prl2013, rosato-criticalinf2008, shao2015local, danziger-ndes2014, boguna-PhysRevX, yanqing-PhysRevX, radicchi-PhysRevX, danziger2015resistors, stochastic2012, amir-naturecomm2012}.
Further studies have also explored more realistic structures such as spatially embedded networks  \cite{bashan-naturephysics2013, wei-prl2012, danziger-jcomnets2014, shekhtman-pre2014,berezin2013spatially, danziger2015two}, and different types of realistic failures such as response under degree-based attacks \cite{dong-pre2013, huang-pre2011} and localized attacks \cite{berezin2013spatially, shao2015local}. Nevertheless, all previous studies on interdependent networks ignored the realistic effect of modularity on the resilience of interdependent networks.

Many real world networks have a modular structure  including biological networks \cite{bullmore2012economy}, infrastructure such as the power grid,  internet \cite{eriksen2003modularity} and airport networks \cite{guimera2005worldwide},  and financial networks \cite{garas2008structural}. Several studies have explored the robustness of individual modular systems (i.e. single networks) \cite{bagrow2011robustness, babaei2011cascading} yet no study has considered the effect of interdependence in modular networks. It is well known that many of these systems are also interdependent and thus it is crucial to understand how the modular structure affects the resilience of interdependent networks. 

There has also been considerable work on understanding various types of attacks on networks \cite{cohen-prl2001,callaway-prl2000, shao2015local, berezin2013spatially, gallos2005stability}. Recent work by Shai et al. \cite{shai2014resilience} developed an analytical method where the attack is carried out on interconnected nodes, i.e. nodes that connect communities. Further work by da Cunha et al. \cite{da2015complex} showed that in real networks, attacks on interconnected nodes are even more damaging than attacks based on betweenness. It is particularly important to consider the effectiveness of this type of attack in interdependent networks since the researchers in \cite{da2015complex} showed that the US power grid is among the most susceptible networks to this type of attack and it is well known that power grids are interdependent with other infrastructures.

In our model we assume modular networks composed of $m$ modules, and a fixed ratio, $\alpha$, between the probability for an intramodule link and intermodule link. We further fix the total average degree, $k_\text{tot}$, of the network. Using these three parameters we can determine the average intramodule degree, $k_\text{intra}$ and the average intermodule degree $k_\text{inter}$. We obtain
\begin{align}
\frac{k_\text{intra}}{k_\text{inter}}&=\frac{\alpha}{m-1} \\
k_\text{tot}&=k_\text{intra}+k_\text{inter}.
\end{align}

\begin{figure*}[htbp]
\centering

\begin{subfigure}[b]{0.33\textwidth}
\centering
\includegraphics[ width=0.9\linewidth]{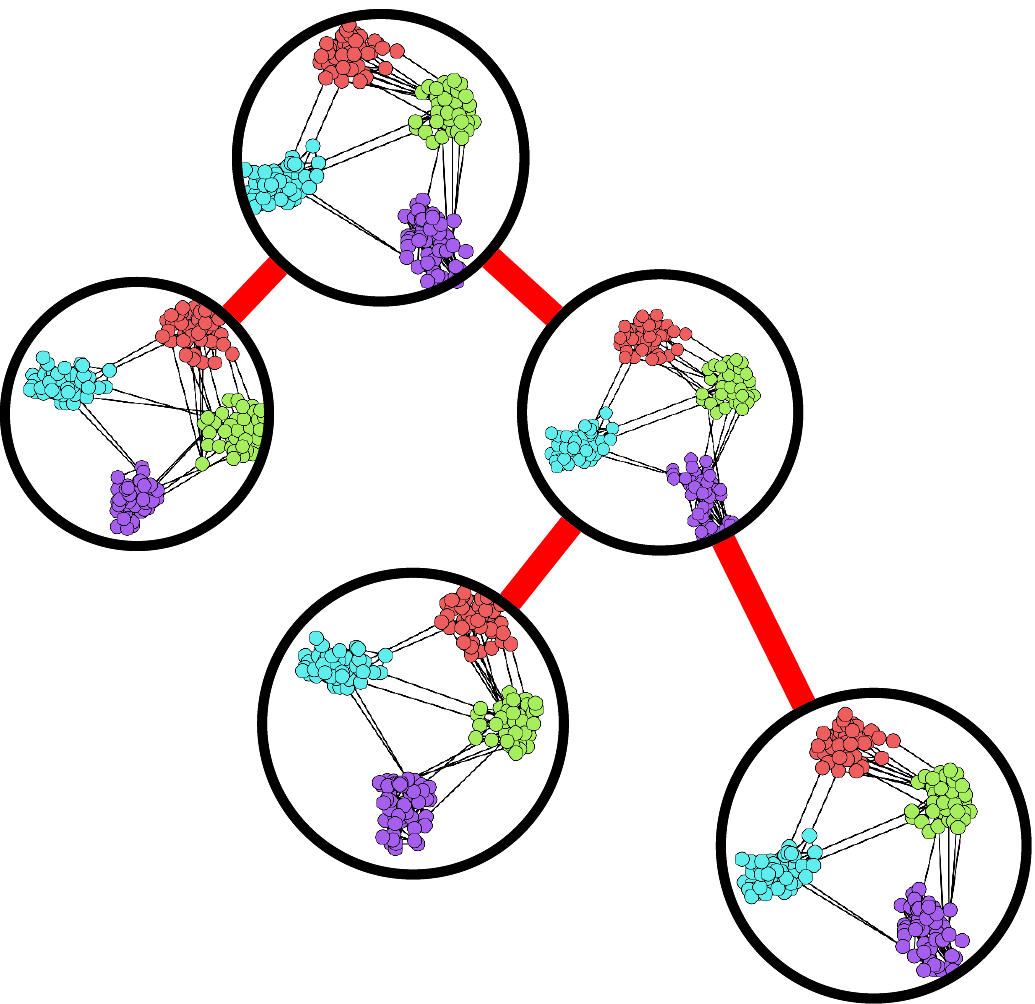} 
\caption{ } 
\label{fig:4mod-tree}
\end{subfigure}
\hfill
\begin{subfigure}[b]{0.33\textwidth}
\centering
\includegraphics[width=0.7\linewidth]{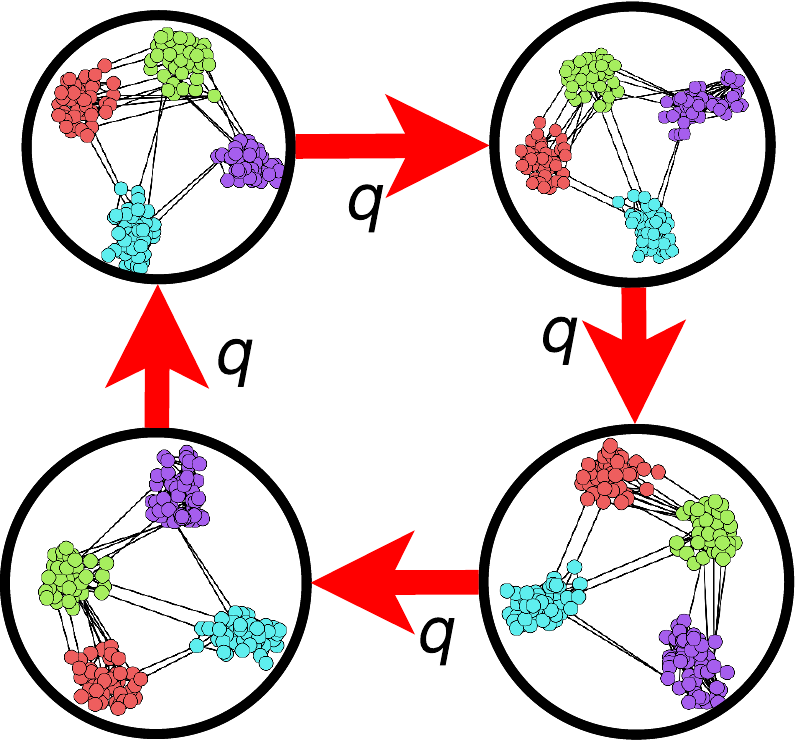} 
\caption{ } 
\label{fig:4mod-loop}
\end{subfigure}
\hfill
\begin{subfigure}[b]{0.32\textwidth}
\centering
\includegraphics[width=1.0\linewidth]{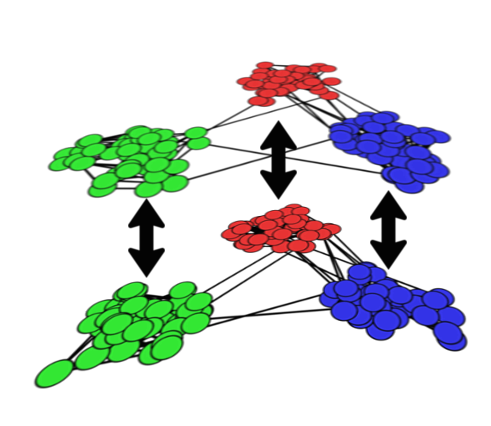} 
\caption{ } 
\label{fig:dep-example}
\end{subfigure}

\caption{ (Color online) We study two possible configurations of a network of networks. ({\bf a}) A treelike network of networks with full coupling and bidirectional dependency links and ({\bf b}) a looplike network of networks with a fraction of dependent nodes, $q$, and unidirectional dependency links. In both ({\bf a}) and ({\bf b}) dependency links are restricted such that they only connect nodes within the same communities,  i.e a node in module $m_a$ in network $i$ will depend on a node also in module $m_a$ in network $j$.  ({\bf c}) Demonstration of the dependency between a pair of interdependent networks shown in ({\bf a}) and ({\bf (b}). The dependency is between the same communities in different networks (same colors). }
\label{fig:diagram-NoN}
\end{figure*}

Note that $k_\text{inter}$ increases with $m$, since networks with more modules have more interlinks \cite{shai2014resilience}. We generate $n$ of these modular networks, and for simplicity we assume that each network has the same $m$, $\alpha$ and $k$. We then create dependency links between the nodes in the different networks.  The fraction of nodes in network $i$ which depend on nodes in network $j$ is defined as $q_{ij}$. The dependency links are either bidirectional (in our analysis of treelike dependencies, Sec. III, Fig. \ref{fig:4mod-tree}) or unidirectional (in our analysis of looplike dependencies, Sec. IV, Fig. \ref{fig:4mod-loop}). In both cases we further restrict the dependency links such that a node in community $m_a$ in network $i$ will depend on a node in community $m_a$ in network $j$, i.e. the dependency links are within the same community. This restriction is particularly reasonable in our example of cities where each city has its own tightly connected, interdependent infrastructure with relatively few connectivity links between the cities. 

We focus on the case of attack on the interconnected nodes, i.e. nodes with at least one connectivity link to a different module. Many studies have shown that these links are particularly susceptible to failure in biological networks and serve as efficient targets in attacking infrastructure \cite{da2015complex}. Further in our example of cities, these nodes have the longer distance links that are more likely to fail \cite{mcandrew2015robustness}.  

Our attack randomly removes a fraction $1-r$, of the interconnected nodes in the network until there are no remaining interconnected nodes and then continues to remove nodes randomly.  The fraction of unremoved interconnected nodes, $r$, is related to the overall fraction of unremoved nodes, $p$, by 
\begin{equation}
r=\frac{p-\bar{p}_{inter}}{1-\bar{p}_{inter}}
\label{eq:r-p-general}
\end{equation}
where $\bar{p}_{inter}$ is the probability that a node is \emph{not} interconnected.

We study this model on both treelike networks of networks (NoNs) and looplike NoNs (Fig. \ref{fig:diagram-NoN}) and find that there are two distinct regimes depending on the number of modules, $m$. In the first regime, for small $m$, the network first separates into isolated, interdependent, yet functional modules, i.e. there is a transition at $r=0$. However, for larger $m$ the network collapses due to the removal of a finite nonzero fraction of interconnected nodes. We provide an analytic solution that predicts the critical point, $m^*$, where the system has a transition from one behavior to the other, as well as solutions for the size of the giant component as a function of the fraction of removed nodes. We support our theory by simulations.

\section{Failure and Attack on Interdependent Modular Networks}
We  now extend the method of Callaway et al. \cite{callaway-prl2000} to the case of interdependent networks or network of networks (NoN). We will then specifically apply this framework to the case of interdependent modular networks. 

We begin by recalling the derivation in \cite{callaway-prl2000} for a single network $i$ with a given degree distribution $P_i(k)$ described by the generating function
\begin{equation}
G(x)=\sum_{k=0}^{\infty}P_i(k)x^k.
\end{equation}
For degree based attacks where nodes with degree $k$ are removed with probability $1-r_k$, the generating function is \cite{callaway-prl2000}
\begin{equation}
F_0(x)=\sum_{k=0}^\infty r_kP_i(k)x^k.
\label{eq:F-single}
\end{equation}
The generating function of the branching process is then 
\begin{equation}
F_1(x)=F_0'(x)/G'(x).
\end{equation}
Following Callaway et al. \cite{callaway-prl2000} we obtain the probability that a randomly chosen edge leads to a cluster of given size, $H_1(x)$, as
\begin{equation}
H_1(x)=1-F_1(x)+xF_1\left[H_1(x)\right]
\label{eq:cluster1}
\end{equation}
and the probability that a randomly chosen node leads to a cluster of a given size, $H_0(x)$, 
\begin{equation}
H_0(x)=1-F_0(1)+xF_0\left[H_1(x)\right].
\label{eq:cluster2}
\end{equation}
The fraction of nodes in the giant component is 
\begin{equation}
P_\infty(x)=1-H_0(1)=F_0(1)-F_0(u),
\label{eq:final-single}
\end{equation}
with $u$ being the smallest non-negative real solution of the self consistency equation
\begin{equation}
u=1-F_1(1)+F_1(u).
\label{eq:u}
\end{equation}

In order to generalize this framework for the case of attack on a NoN we must include the fact that nodes have an additional uniform likelihood to fail, $1-p_\text{dep}$, where $p_\text{dep}$ comes from the effect of the dependency links and depends on the specific topology of the NoN. 

In such a case we have that
\begin{equation}
\tilde{F}_0(x)=p_\text{dep}\sum_{k=0}r_kP_i(k)x^k
\label{eq:NoN-gen}
\end{equation}
where $\tilde{F}_0(x)$ is the anologue of Eq. (\ref{eq:F-single}) for interdependent networks and $p_\text{dep}$ can be extracted from the sum since it does not depend on the degree of the node. The rest of the derivation continues the same for Eqs. (\ref{eq:F-single})-(\ref{eq:final-single}). 

We note that Eq. (\ref{eq:u}) is the same as before only now with the extra factor of $p_\text{dep}$, giving
\begin{equation}
1=p_\text{dep}\frac{F_1(u)-F_1(1)}{u-1},
\label{eq:NoN-final}
\end{equation}
and Eq. (\ref{eq:final-single}) also will have the same factor $p_\text{dep}$. This pair of equations, Eqs. (\ref{eq:final-single}) and (\ref{eq:NoN-final}), can be used to find the giant component of a NoN under any sort of degree based attack and can be used as an alternative method to that of Huang et al. \cite{huang-pre2011} and Dong et al. \cite{dong-pre2013}.

We will now generalize this framework to the case where each network in the NoN is a modular network with the parameters $m$, $\alpha$, and $k$, defined above. 
For the case of a modular network where the modules are made up of Erd\H{o}s-R\'enyi structures, Eq. (\ref{eq:r-p-general}) becomes
\begin{equation}
r=\frac{p-e^{-k_\text{inter}}}{1-e^{-k_\text{inter}}}.
\label{eq:r-p}
\end{equation}
We recall the results from Shai et al. \cite{shai2014resilience} for the generating functions
\begin{align}
G(x)&=e^{-(k_{intra}+k_{inter})(x-1)} \\
F_0(x)&=e^{k_{intra}(x-1)-k_{inter}}(1-r)+rG(x) \\
F_1(x)&=F'_0(x)/G'(x).
\end{align}
We now combine the results from Shai et al. \cite{shai2014resilience} for a single network with Eqs. (\ref{eq:final-single}) and (\ref{eq:NoN-final}) for a NoN. We note that these equations not only describe the case of an interdependent modular NoN but also the case where we have both random failure with probability $1-p_\text{rand}$ and targeted attack. For our purposes we will assume that $p_\text{rand}=p_\text{dep}$, i.e. the random damage is solely the result of the dependencies. Our new generating functions are thus,
\begin{align}
& \tilde{F}_0(x) = p_\text{dep} F_0(x) \nonumber \\
& \tilde{F}_{1}(x) = p_\text{dep} F_{1}(x).
\end{align}
The average connected component size is given by Leicht and D'Souza \cite{leichtdsouza2009},
\begin{equation}
\langle s \rangle =p_\text{dep} F_0(1)+p_\text{dep}rk_\text{intra}F_0(1)j_0+p_\text{dep}rk_\text{inter} j_1
\end{equation}
where
\begin{align}
j_0&=p_\text{dep}F_0(1)+p_\text{dep}rk_\text{intra}F_0(1)j_0+p_\text{dep}rk_\text{inter}j_1 \\
j_1&=p_\text{dep}r+p_\text{dep}rk_\text{intra}j_0+p_\text{dep}rk_\text{inter}j_1.
\end{align}

Combining the above equations, we find that there is a giant component that spans the system when
\begin{widetext}
\begin{align}
& \big( 1 - p_\text{rand}k_\text{intra} (e^{-k_\text{inter}} + r_c- r_c e^{-k_\text{inter}} ) \big) (1 - p_\text{rand} r_c k_\text{inter}) - p_\text{rand}^2 r_c^2 k_\text{intra} k_\text{inter} = 0 \nonumber \\
& \Rightarrow r_c^2 \big(  p_\text{rand}^2 k_\text{intra} k_\text{inter} e^{-k_\text{inter}}   \big) +  r_c  \big( p_\text{rand} k_\text{inter} + p_\text{rand} k_\text{intra} - p_\text{rand} k_\text{intra} e^{-k_\text{inter}} - p_\text{rand}^2 k_\text{intra} k_\text{inter} e^{-k_\text{inter}} \big) + \big( p_\text{rand} k_\text{intra} e^{-k_\text{inter}} -1 \big) = 0
\label{eq:r-c}
\end{align}
\end{widetext}
which can be solved using the quadratic formula for $r_c$.

\section{Treelike network of interdependent modular networks }

We now consider the case of a treelike NoN (see Fig. \ref{fig:4mod-tree}) with full dependency and no-feedback, i.e.  if network $i$ depends on network $j$ then each node in $i$ depends on a single node in $j$ and vice versa.  As stated earlier, this can serve as a good model of infrastructure across cities. We remove a fraction $1-r$ of the interconnected nodes from one of the networks and aim to obtain the mutual giant connected component. The fraction of nodes which fail due to the effect of the dependencies is given by 
\begin{equation}
p_\text{dep}=(1-e^{-(k_\text{intra}+k_\text{inter})P_\infty})^{n-1}.
\label{eq:dep-trees}
\end{equation}
This can be understood by noting that each node is a part of a $n$-tuple of interdependent nodes and we require that all these $n$ nodes be in their networks respective giant components. Therefore, besides the node itself, there are $n-1$ dependent nodes that must all be in their networks' respective giant components after the initial attack.

\begin{figure*}
\centering
\hfill
\begin{subfigure}{0.49\textwidth}
\centering
\includegraphics[width=1.0\linewidth]{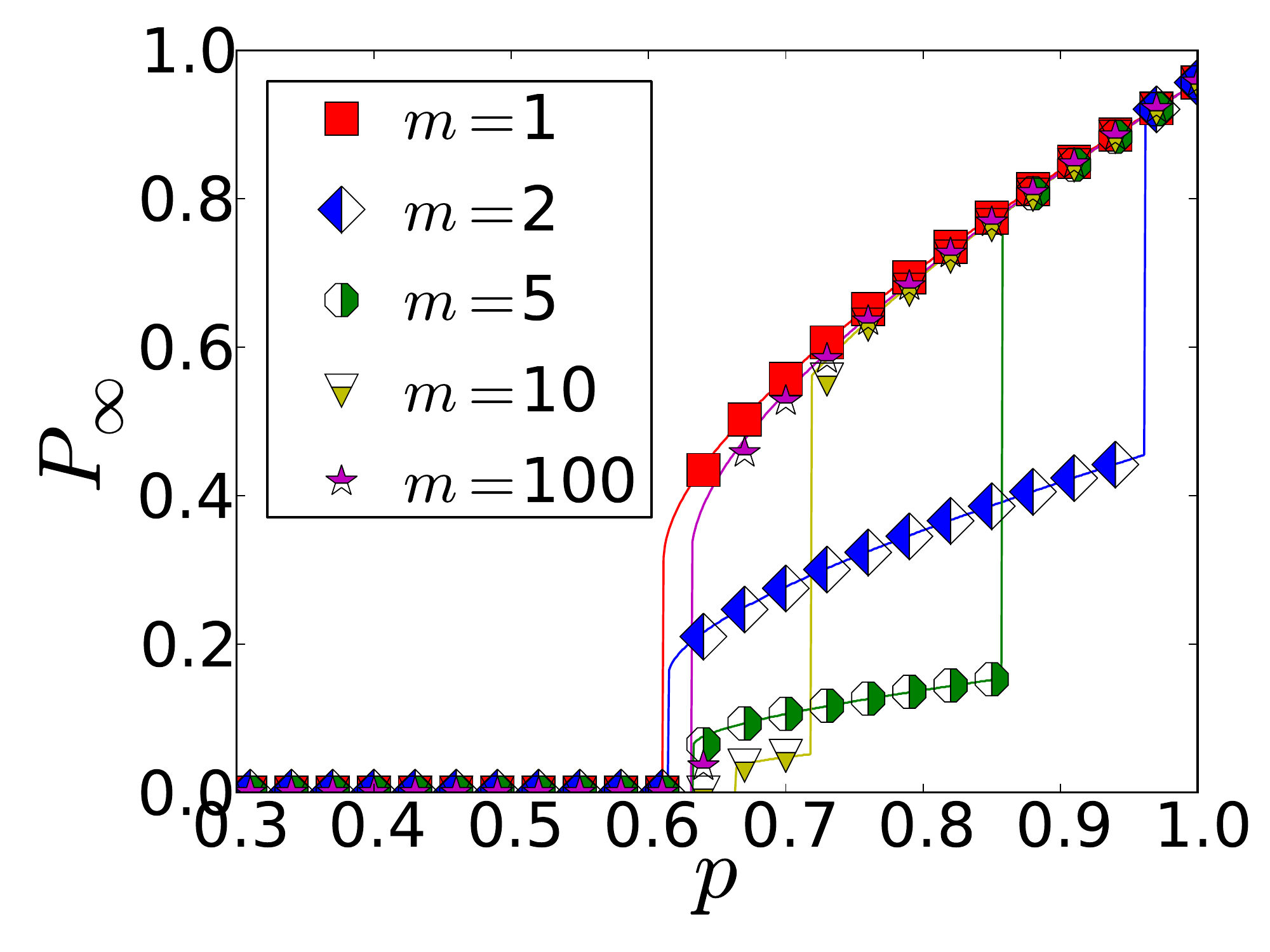} 
\caption{ } 
\label{fig:modular-intra-dep}
\end{subfigure}
\hfill
\begin{subfigure}{0.49\textwidth}
\centering
\includegraphics[width=1.0\linewidth]{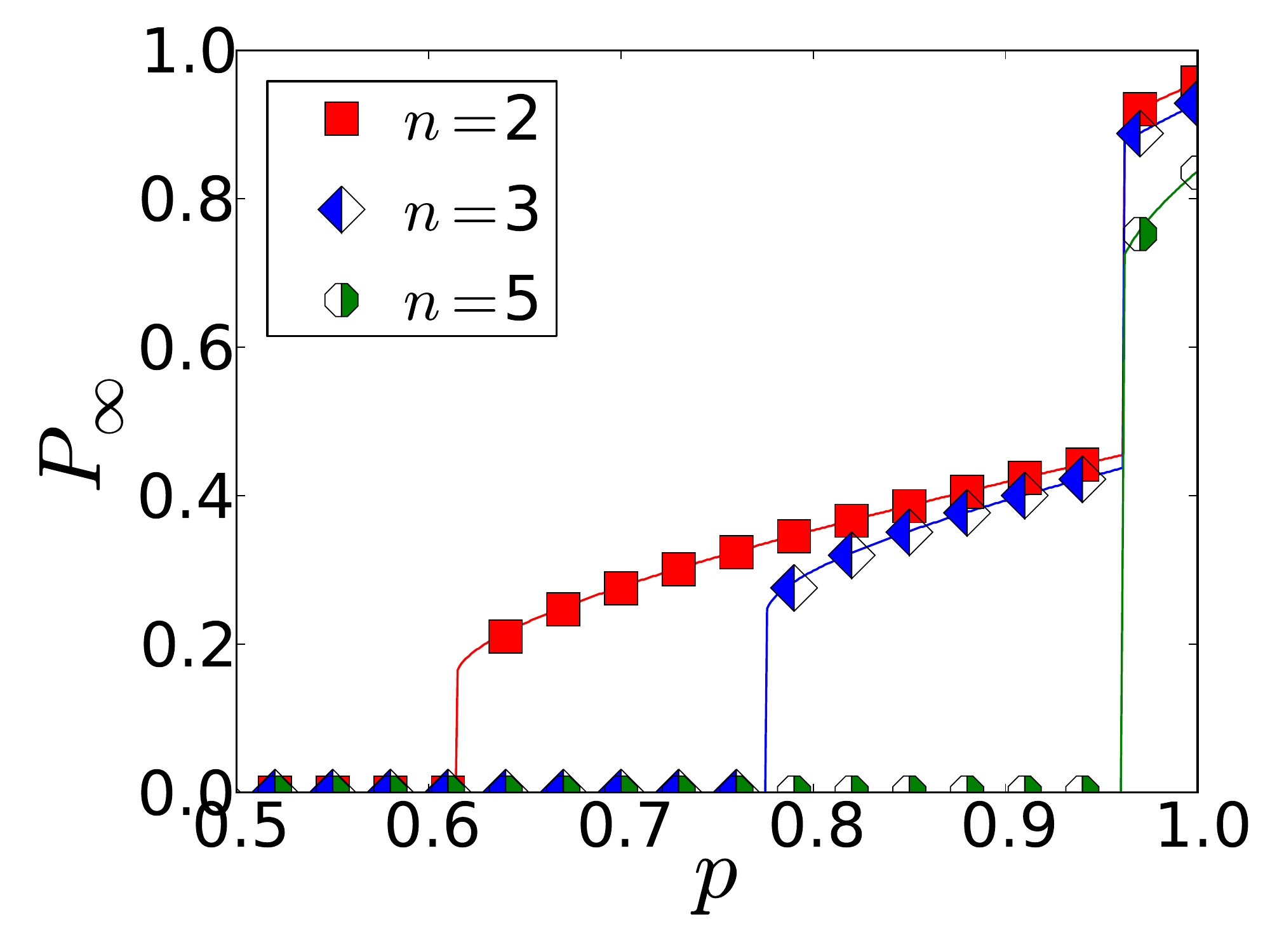} 
\caption{ }
\label{fig:modular-intra-dep-nnets}
\end{subfigure}
\vfill
\centering
\begin{subfigure}{0.49\textwidth}
\centering
\includegraphics[width=1.0\linewidth]{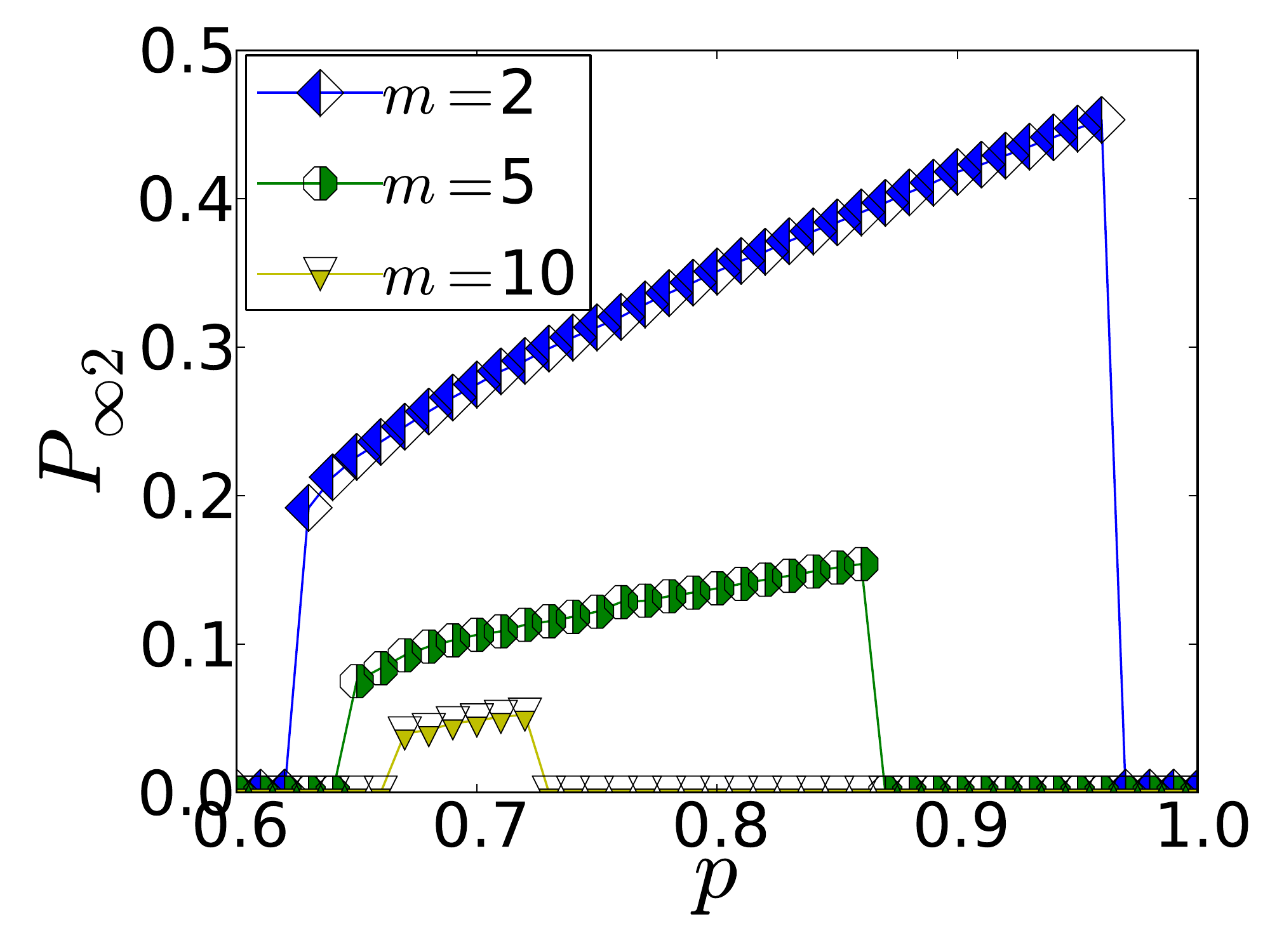} 
\caption{}
\label{fig:gcc2}
\end{subfigure}
\hfill
\begin{subfigure}{0.49\linewidth}
\centering
\includegraphics[width=1.0\linewidth]{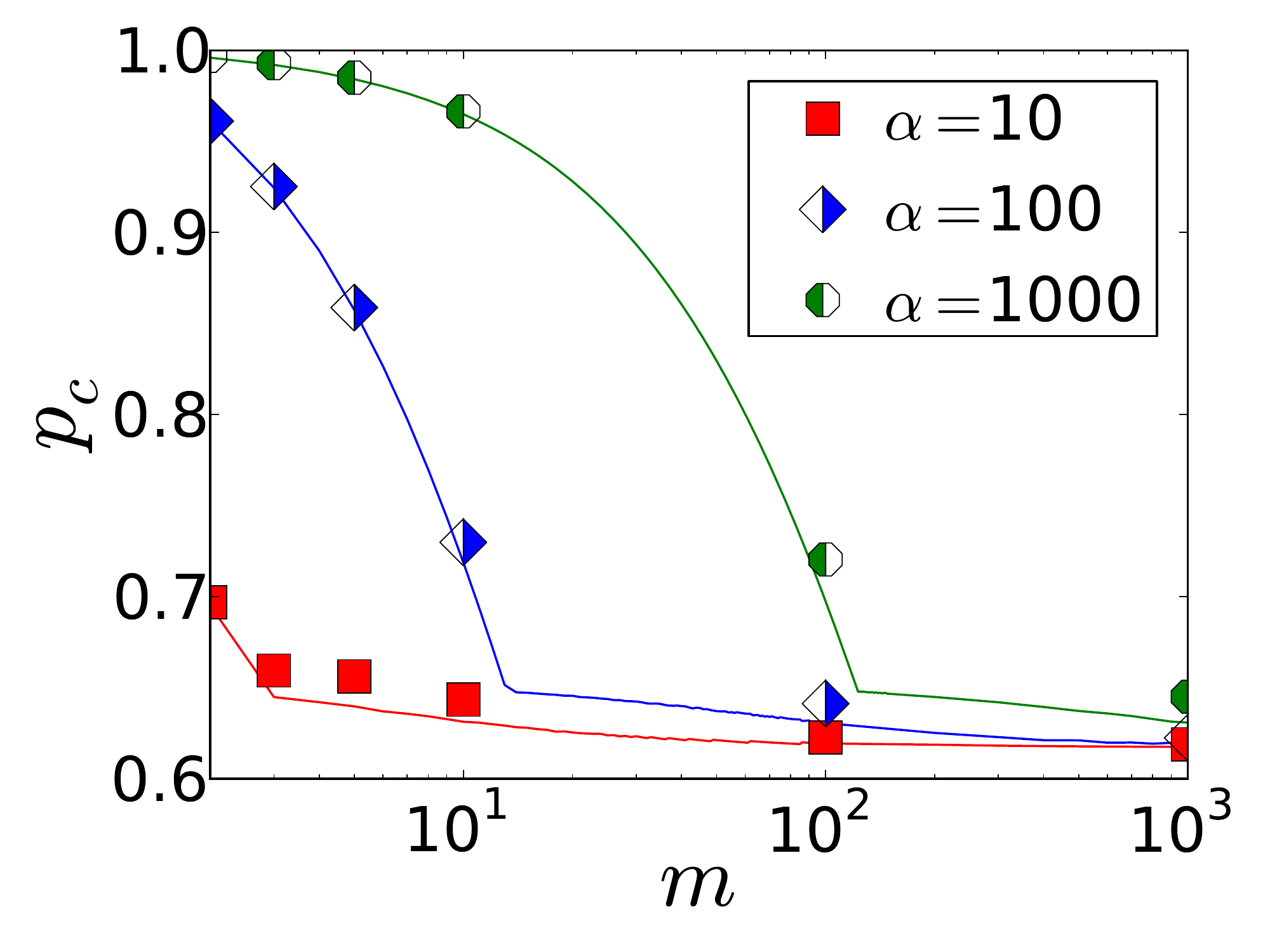} 
\caption{}
\label{fig:pc1}
\end{subfigure}
\hfill
\caption{{\bf (a)} and {\bf (b)} Results of simulations and theory according to Eq. (\ref{eq:mod_intra-dep}) for modular networks with $\alpha=100$ and $k_\text{tot}=4$. In {\bf (a)} we vary $m$ with $n=2$  and in {\bf (b)} we vary $n$ with $m=2$. Note that by varying either of these parameters we may have either one or two first order transitions (at two different values of $p$). Note that for the limiting case, $m=1$ and $n=2$, i.e. the squares in {\bf (a)} we obtain the known result of Buldyrev et al. \cite{buldyrev-nature2010} $p_c=2.445/k_{tot}=0.611$ for $k_{tot}=4$. For larger $m$, $p_c$ increases, implying that modular networks are more vulnerable under attack. In {\bf (c)} we show the size of the second largest giant  component, $P_{\infty2}$, for the same parameters as in {\bf (a)}. This plot verifies that the first abrupt drop in $P_\infty$ is the result of modules separating and there is a regime where the largest and second largest components are essentially equivalent. Note as the number of modules increases the size of the second largest component decreases, yet there are more surviving components, i.e. for $m=10$ there are $10$ modules of similar size. In {\bf (d)} we show $p_c$, defined as the point of the first abrupt drop (at the largest $p$), as a function of $m$ for several values of $\alpha$ with $k_\text{tot}=4$. Note that below $m^*$ $p_c$ drops quickly according to Eq. (\ref{eq:pc_r}) yet as $m$ increases it decreases slowly according to Eq. (\ref{eq:pc-tree-mod}). The sharp kink where $p_c$ changes behaviors is $m^*$ and for $m>m^*$ there is only one single transition rather than two. }
\label{fig:modular-intra-dep-all}
\end{figure*}

\begin{figure*}[htbp]
\centering
\begin{subfigure}{0.45\textwidth}
\centering
\includegraphics[ width=0.9\linewidth]{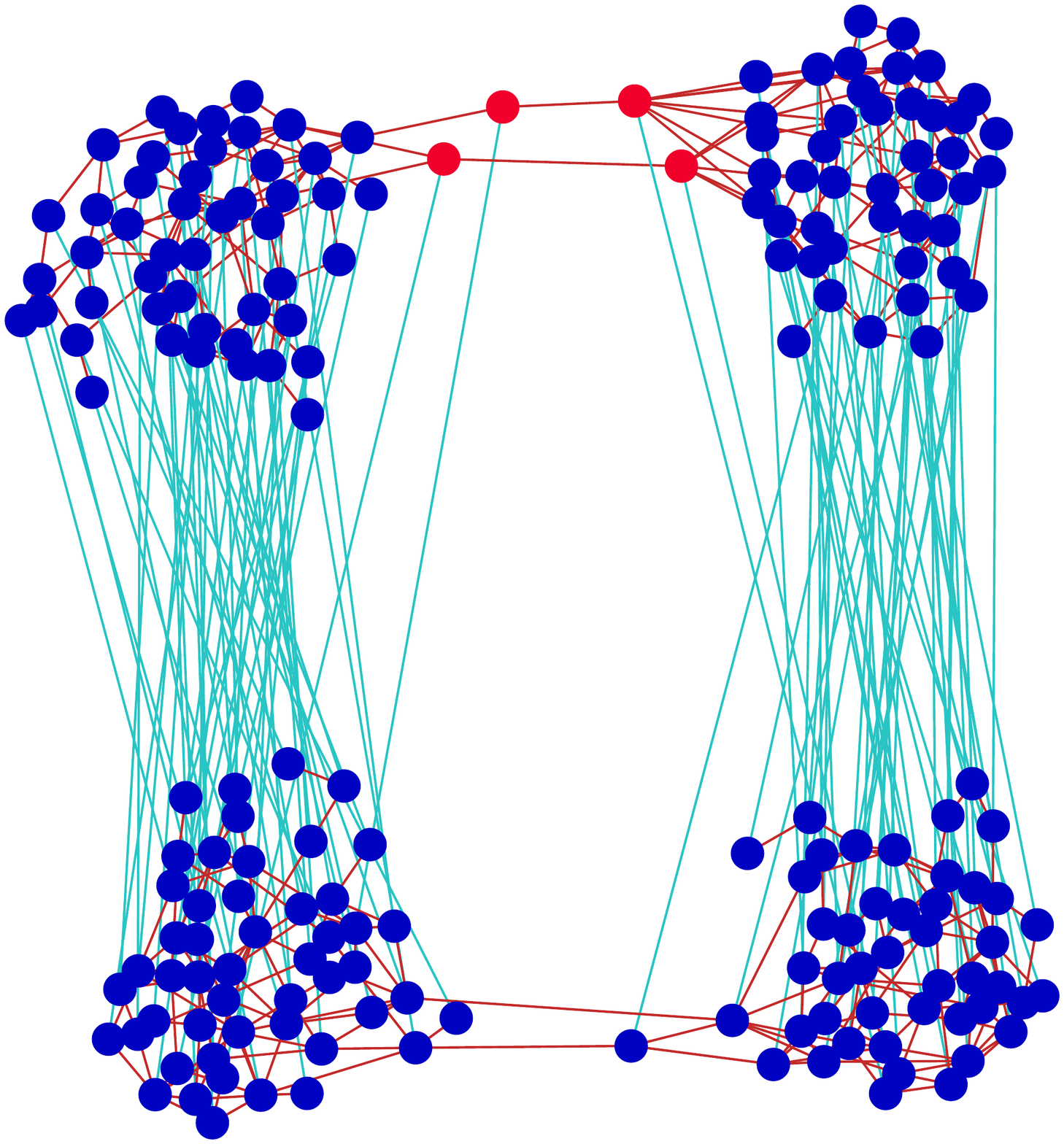} 
\caption{ } 
\label{fig:init-state}
\end{subfigure}
\hfill
\begin{subfigure}{0.45\textwidth}
\centering
\includegraphics[width=0.9\linewidth]{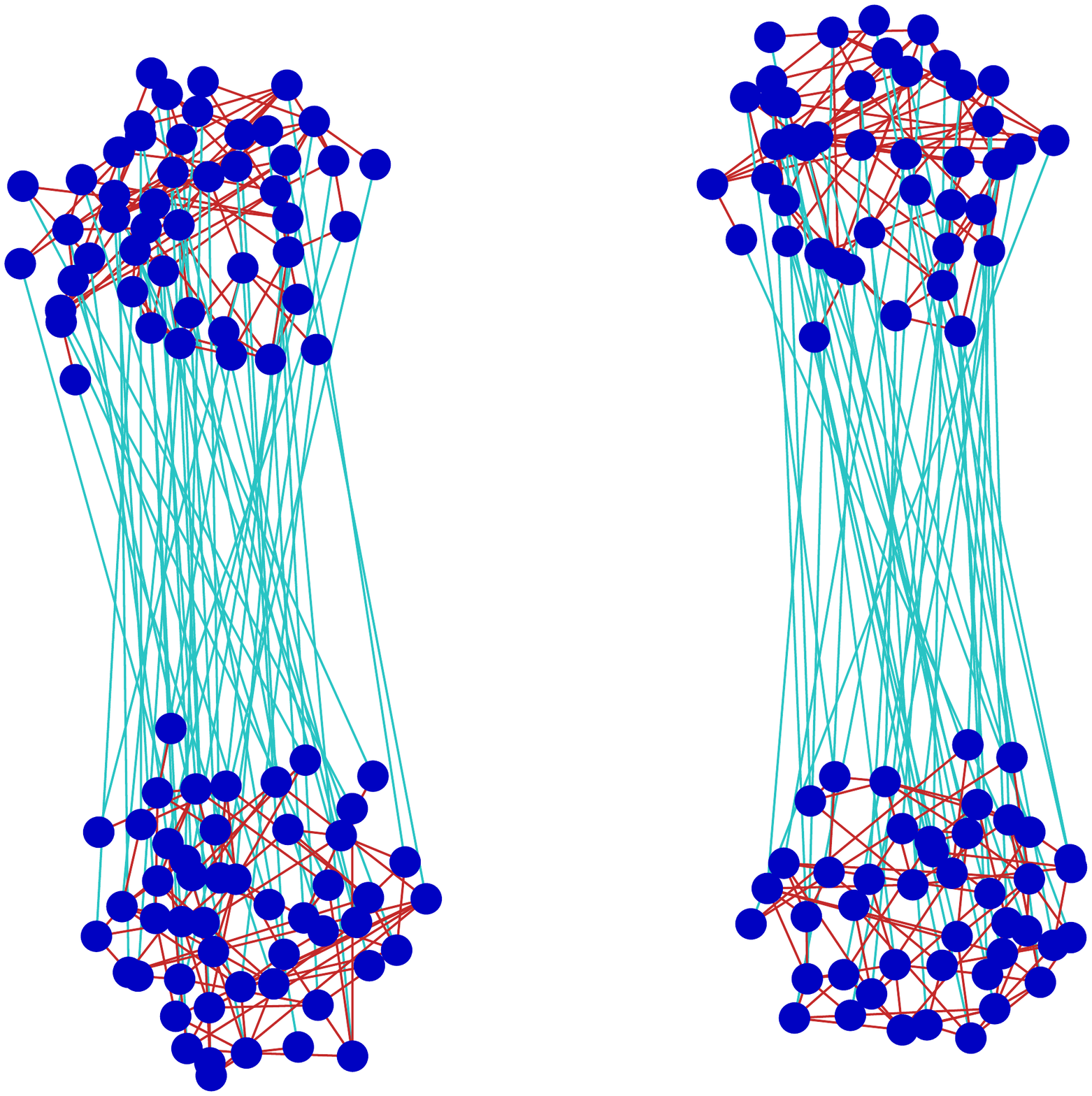} 
\caption{ }
\label{fig:after-separation}
\end{subfigure}

\caption{{\bf a. }Demonstration of the network in the initial state before the attack. The red links are connectivity links and the torquoise links are dependency links. We highlight the interconnected nodes which will be removed by the attack (note that we only perform the attack on one of the networks). {\bf b.} Demonstration of the network after the removal of the interconnected nodes. As shown, the attack divides both networks into their separate modules and we have two components each of which behaves like a typical interdependent network of Erd\H{o}s-R\'enyi networks.}
\label{fig:pic}
\end{figure*}

If we combine Eqs.  (\ref{eq:NoN-gen}), (\ref{eq:NoN-final}), and (\ref{eq:dep-trees}) we obtain the mutual giant connected component in the NoN,
\begin{widetext}  
\begin{equation}
P_\infty=
\begin{cases} 
\Big ( e^{-k_\text{inter}}(1-r)(1-e^{-k_\text{intra}P_\infty})+r(1-e^{-(k_\text{intra}+k_\text{inter})P_\infty})\Big )  (1-e^{-(k_\text{intra}+k_\text{inter})P_\infty})^{n-1} & 0<r<1 \\
\tfrac{p}{m}(1-e^{-mk_\text{intra}P_\infty})^{n} & r<0.
\end{cases}
\label{eq:mod_intra-dep}
\end{equation}
\end{widetext}
Since the dependency links are within modules, Eq. (\ref{eq:mod_intra-dep}) for $r<0$ is simply the equation for a treelike NoN with $k=k_\text{intra}$ and a giant component that is smaller by a factor of $m$. Nonetheless the value of $p_c$ will be the same as for a regular treelike NoN with $k=k_\text{intra}$. We show analytic solutions of these equations along with simulations in Fig. \ref{fig:modular-intra-dep-all} for varying $m$ and varying $n$. By varying either of these parameters we may have either one or two abrupt percolation transitions. Note that for the case of varying $n$ the point of the first abrupt drop is the same for all curves. This is because the point where the modules separate depends only on $m$, $\alpha$ and $k_\text{tot}$, but not on $n$ (see Fig. \ref{fig:pic}). It is worth noting though that for larger $n$, the second regime, where modules exist independently (see Fig. \ref{fig:after-separation}), may not be present if the individual modules do not have enough intralinks to survive the damage from the dependencies.

We demonstrate that in the case where there are two abrupt transitions, the first transition (at higher $p$) represents the separation of the modules and we therefore have $m$ individual modules functioning separately. This is indicated by the spike in the size of the second largest component, $P_{\infty2}$, in Fig. ~\ref{fig:gcc2} and is demonstrated visually in Fig. \ref{fig:pic}. In general for a network composed of $m$ modules, there are $m$ equal components functioning independently. After the separation into modules, we observe a second abrupt transition due to the interdependence.

As $m$ increases, the network begins to act more like a regular network of $n$ interdependent Erd\H{o}s-R\'enyi networks. It is clear from the framework developed until now that there exists a critical value of $m$, $m^*$ above which the system undergoes a single transition whereas below $m^*$ the system first separates into separate modules before collapsing entirely.

We will now determine $p_c$ and $m^*$ for this system using the framework developed in Chap. II. For certain regimes there exist two first-order, discontinuous jumps in the size of the giant component (see Fig. \ref{fig:modular-intra-dep}). We define $p_c$ as the point where the first (higher $p$) jump occurs, regardless of whether there is a second jump afterwards. If desired, the point of the second jump can be found by solving for the critical threshold for a typical Erd\H{o}s-R\'enyi  NoN with $k=k_\text{intra}$.  We will now solve for $p_c$ by first finding $r_c$ using Eq. (\ref{eq:r-c}) and then converting it to $p_c$ using Eq. (\ref{eq:r-p}). We recall that for the case of trees we use $p_\text{dep}=(1-e^{-(k_\text{intra}+k_\text{inter})P_\infty})^{n-1}$ in Eq. (\ref{eq:r-c}) as was done in Eq. (\ref{eq:mod_intra-dep}). This gives us the following system of equations
\begin{widetext}
\begin{align}
p_\text{dep}&=(1-e^{-(k_\text{intra}+k_\text{inter})P_\infty})^{n-1} \\
P_\infty&= \Big ( e^{-k_\text{inter}}(1-r_c)(1-e^{-k_\text{intra}P_\infty})+r_c(1-e^{-(k_\text{intra}+k_\text{inter})P_\infty})\Big )  (1-e^{-(k_\text{intra}+k_\text{inter})P_\infty})^{n-1} \\
r_c&=\frac{-b+\sqrt{b^2-4ac}}{2a} 
\label{eq:pc-tree-mod}
\end{align}
\end{widetext}
with the values of $a$, $b$ and $c$ being the coefficients from Eq. (\ref{eq:r-c}).

We see in Fig. \ref{fig:pc1} that the simulations fit the theory well except near the point where we switch to having a single abrupt transition. This small discrepancy is because there are large fluctuations in the fraction of individual modules that are part of the giant component, which are not considered in the mean-field approach. Once we move away from this critical point the theory again fits the simulations.

In order to find $m^*$ we compare the value of $p$ when $r=0$ to the $p_c$ for a treelike network of networks with $k=k_\text{intra}$. Essentially this means comparing the fraction of nodes needed to remove all interconnected nodes and the value of $p_c$ where the modules themselves break apart.

We can trivially find the value of $p$ when $r=0$ by noting that $r=\tfrac{p-e^{-k_\text{inter}}}{1-e^{-k_\text{inter}}}$. We obtain in terms of $m$
\begin{equation}
p(r=0)=e^{-k_\text{tot}\tfrac{m-1}{\alpha+m-1}}.
\label{eq:pc_r}
\end{equation}
We note that the value of $p_c$ for a treelike network of networks (without communities) was determined by Gao et al. \cite{gao-prl2011}. In order to find $p_c$ we first find the value of $\bar{k}_{min}$, the  degree at which the system undergoes a transition. This is done using the system of equations
\begin{align}
f_c=-[nW(-\frac{1}{n}e^{-1/n}]^{-1} \\
\bar{k}_{min}=[nf_c(1-f_c)^{(n-1)}]^{-1}
\end{align}
where $W(x)$ is the Lambert function.

Then $p_c=\bar{k}_{min}/<k>$ where $<k>$ is the average degree of each network, in our case $<k>=k_\text{intra}$. We obtain 
\begin{equation}
p_c=\tfrac{\bar{k}_{min}(\alpha+m-1)}{k_\text{tot}\alpha}.
\label{eq:pc1}
\end{equation}

\begin{figure*}
\centering
\hfill
\begin{subfigure}{0.45\textwidth}
\centering
\includegraphics[width=1.0\linewidth]{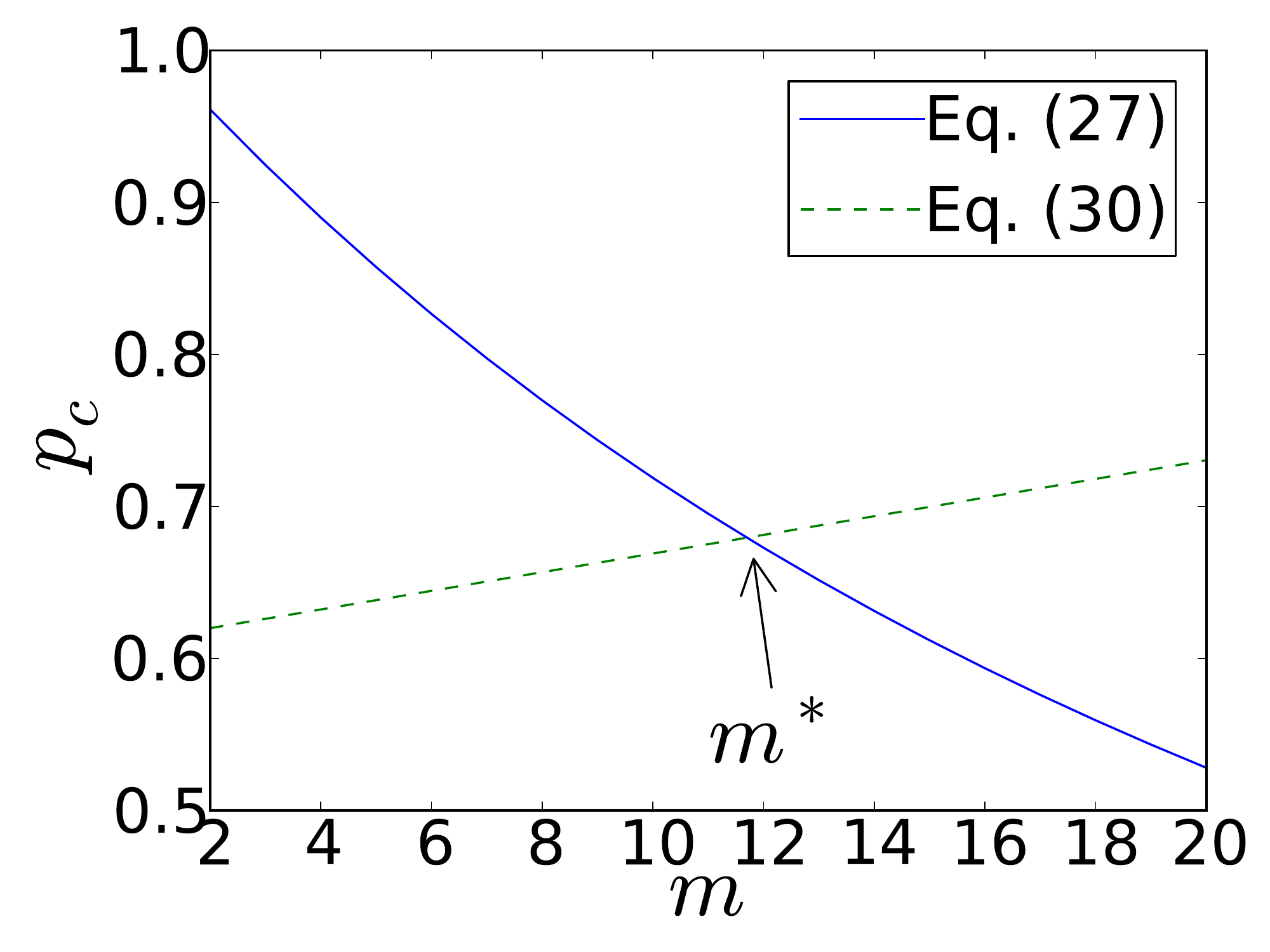} 
\caption{}
\label{fig:m*_det}
\end{subfigure}
\hfill
\begin{subfigure}{0.45\textwidth}
\centering
\includegraphics[width=1.0\linewidth]{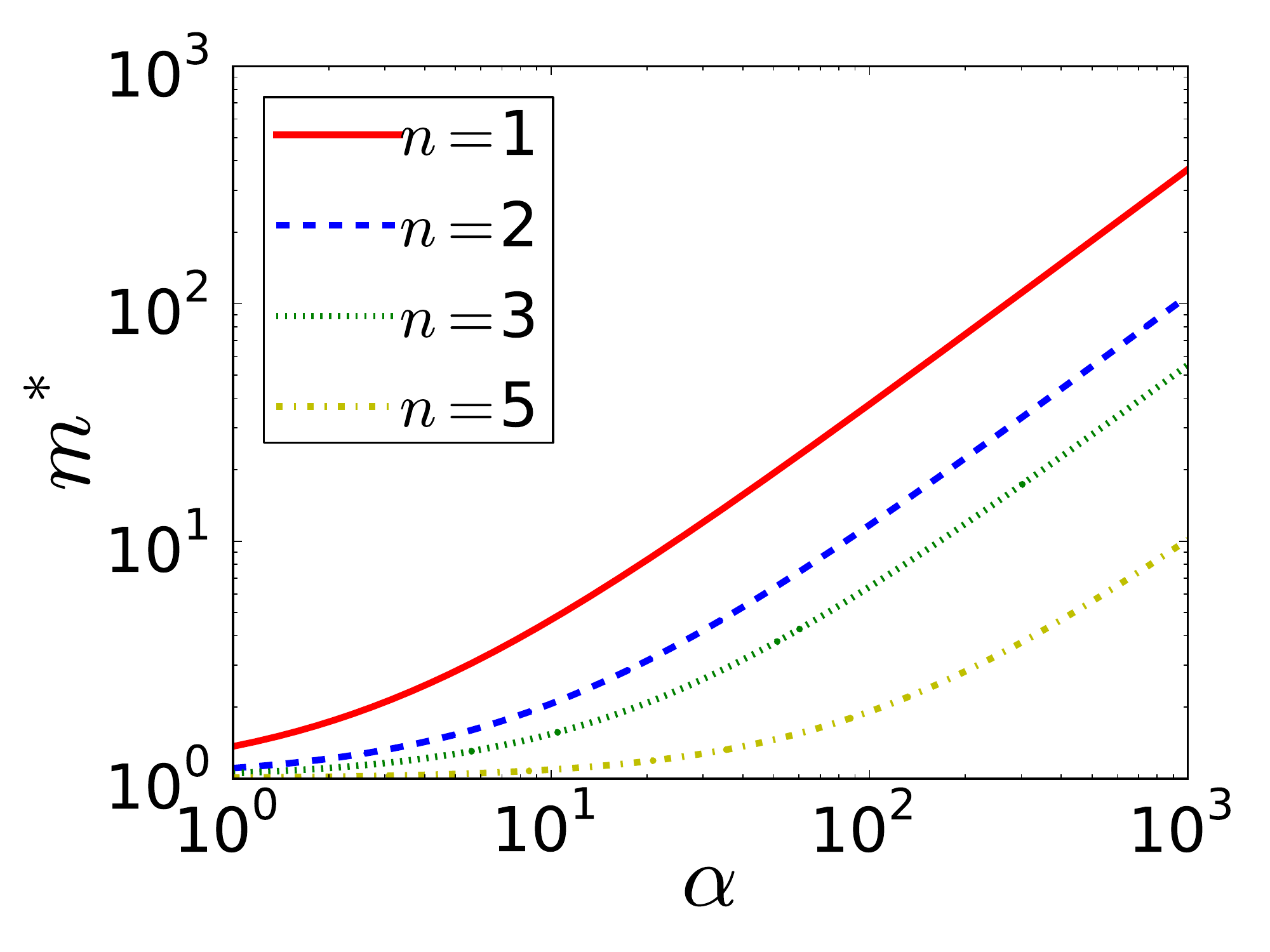} 
\caption{}
\label{fig:m*_vals}
\end{subfigure}
\hfill
\caption{{\bf a.} We show how to obtain $m^*$ by plotting Eq. (\ref{eq:pc_r}) and Eq. (\ref{eq:pc1}) for the case of $n=2$, $k=4$, and $\alpha=100$. The curve showing Eq. (\ref{eq:pc_r}) decreases with $m$ since the number of interlinks increases with $m$ so a larger fraction of nodes have an intermodule connection. On the other hand, $p_c$ in Eq. (\ref{eq:pc1}) increases since for fixed $k$, as $m$ increases $k_{intra}$ decreases so the modules themselves are more vulnerable. The value of $m^*$ is determined by the intersection of the two curves. {\bf b.} The value of $m^*$ vs. $\alpha$ according to Eq. (\ref{eq:m*-trees}) for several values of $n$ with $k=4$ is shown. }
\end{figure*}

The value of $m^*$ is found by setting Eq. (\ref{eq:pc_r}) equal to Eq. (\ref{eq:pc1}) to obtain
\begin{equation}
\tfrac{\bar{k}_{min}(\alpha+m^*-1)}{k_\text{tot}\alpha}=e^{-k_\text{tot}   \tfrac{m^*-1}{\alpha+m^*-1}}.
\label{eq:m*-trees}
\end{equation}
This equation can be solved numerically for any value of $n$ and $\alpha$. Essentially, there are two competing effects in this equation. The left side represents the value of $p_c$ for an interdependent network of networks with $k=k_\text{intra}$ and the right side represents the point where we have removed all nodes with an interlink. As $m$ increases, $k_\text{intra}$ decreases, meaning that the modules become more vulnerable and $p_c$ from the left side of the equation increases. On the other hand, as $m$ increases, $k_\text{inter}$ increases so more nodes have an interlink and $p_c$ on the right side of the equation decreases (more resilient).  We demonstrate these two effects and show the point where they intersect in Fig. \ref{fig:m*_det}. Further we show the value of $m^*$ for various values of $\alpha$ and $n$ in Fig. \ref{fig:m*_vals}. Also, as $n$ increases, $\bar{k}_{min}$ increases, meaning the affect of the interdependence is stronger. Essentially it causes the dashed curve in Fig. \ref{fig:m*_det} to increase, thus decreasing $m*$, the point where the curves intersect. Thus for larger $n$ we have a lower $m^*$, as seen in Fig. \ref{fig:m*_vals}.

\section{ Loop of interdependent modular networks}
We will now demonstrate our approach on another example where we consider the specific case of a directed loop of modular networks. In this case, for simplicity, we perform the initial attack on all of the networks rather than just one of them. Also, whereas the dependency links in Chap III. were bidirectional here we use unidirectional links as part of a loop (see Fig. \ref{fig:4mod-loop}). The fraction of nodes which are interdependent is defined as $q$, i.e. $1-q$ is the fraction of autonomous nodes, and is assumed here, for simplicity, to be the same for all pairs of networks. Here again, we will only consider dependency links which are restricted to being between two networks but within the same module. It has been shown that for looplike NoNs the number of networks in the loop is not relevant to the calculation \cite{gao-naturephysics2012, gao-general-net, shekhtman-pre2014}.

\begin{figure*}
\centering
\hfill
\begin{subfigure}{0.49\textwidth}
\centering
\includegraphics[width=1.0\linewidth]{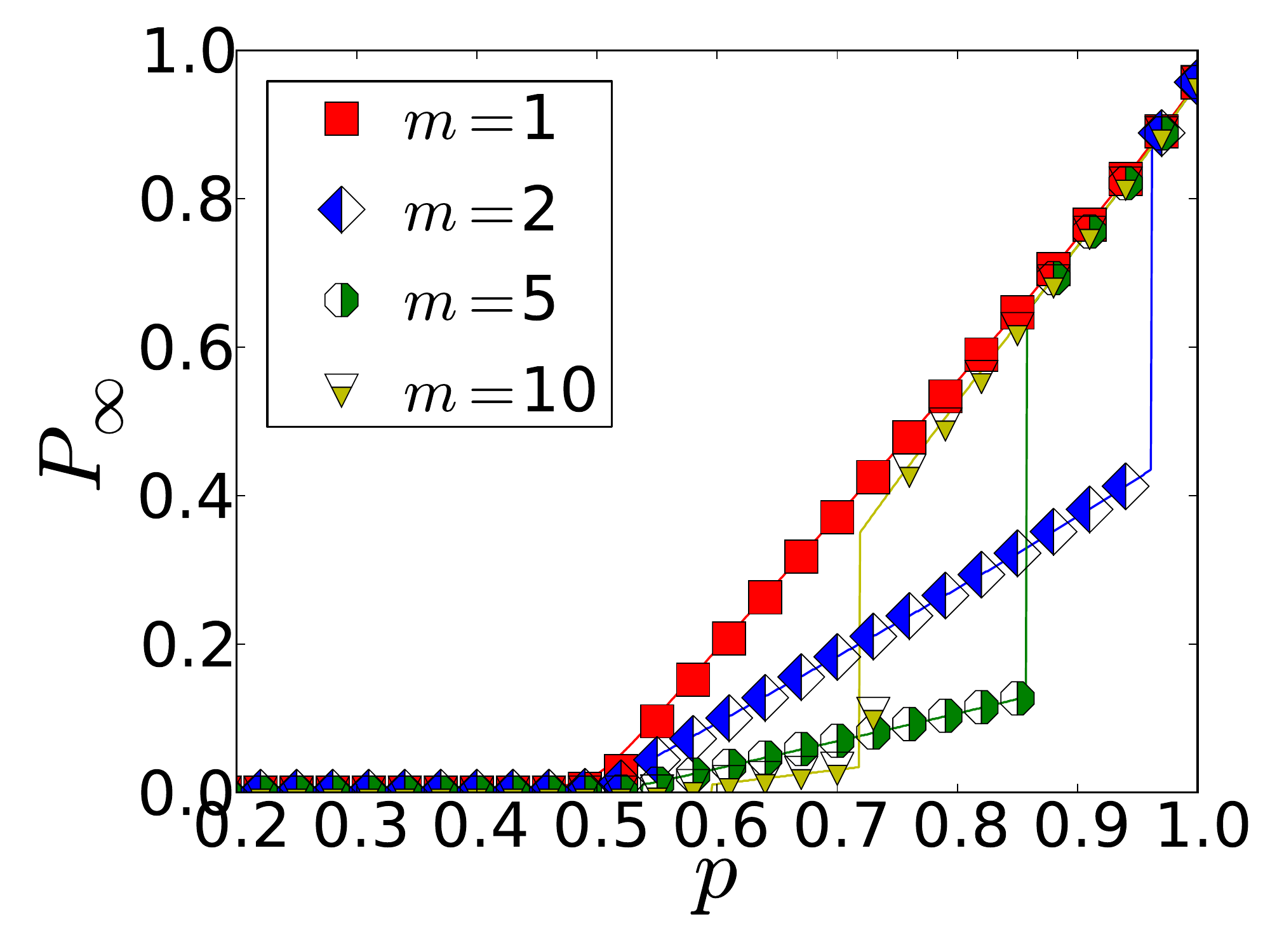} 
\caption{}
\label{fig:RR-varym}
\end{subfigure}
\hfill
\begin{subfigure}{0.49\textwidth}
\centering
\includegraphics[width=1.0\linewidth]{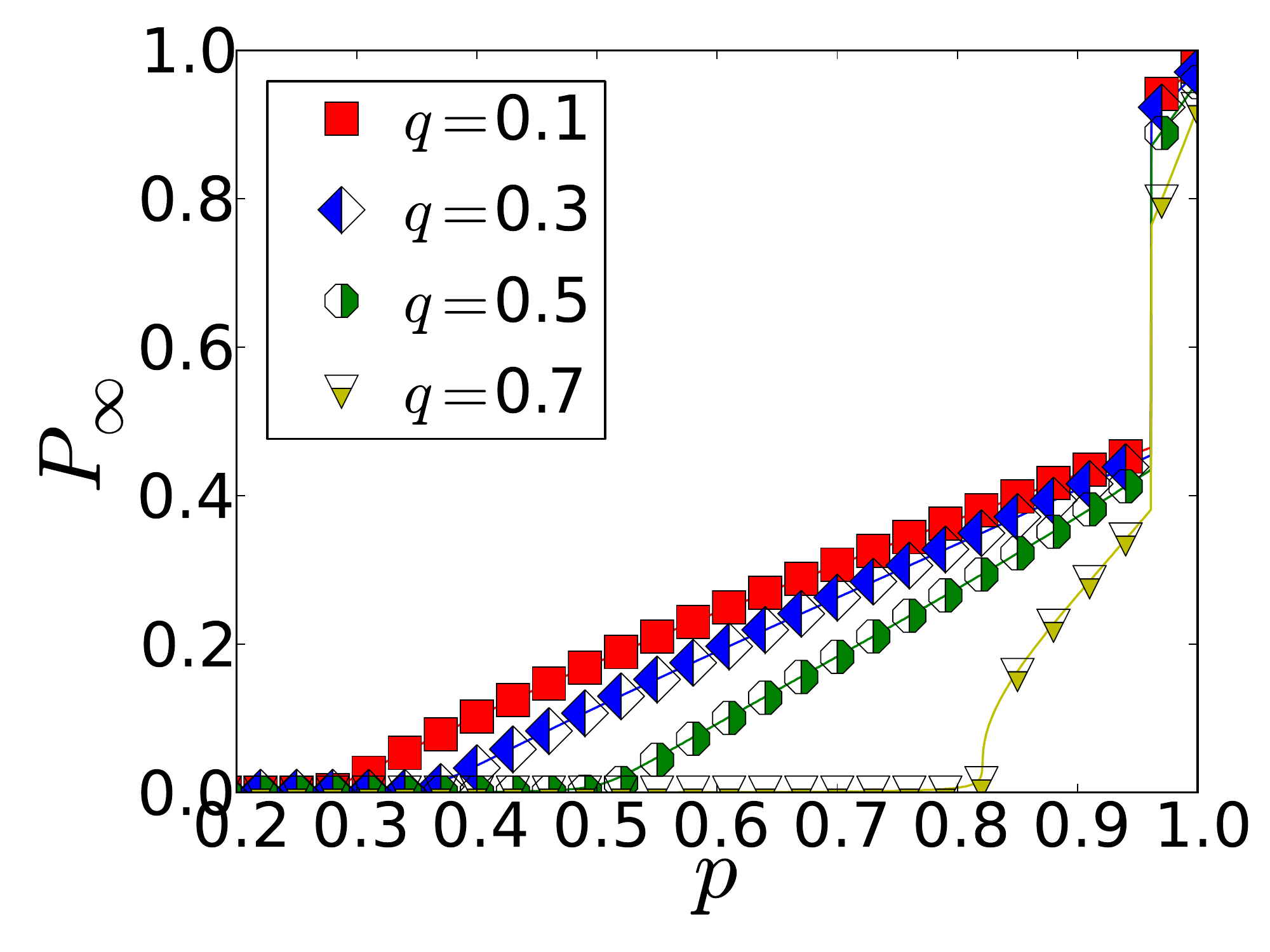} 
\caption{}
\label{fig:RR-varyq}
\end{subfigure}
\vfill

\begin{subfigure}{0.45\textwidth}
\centering
\includegraphics[width=1.0\linewidth]{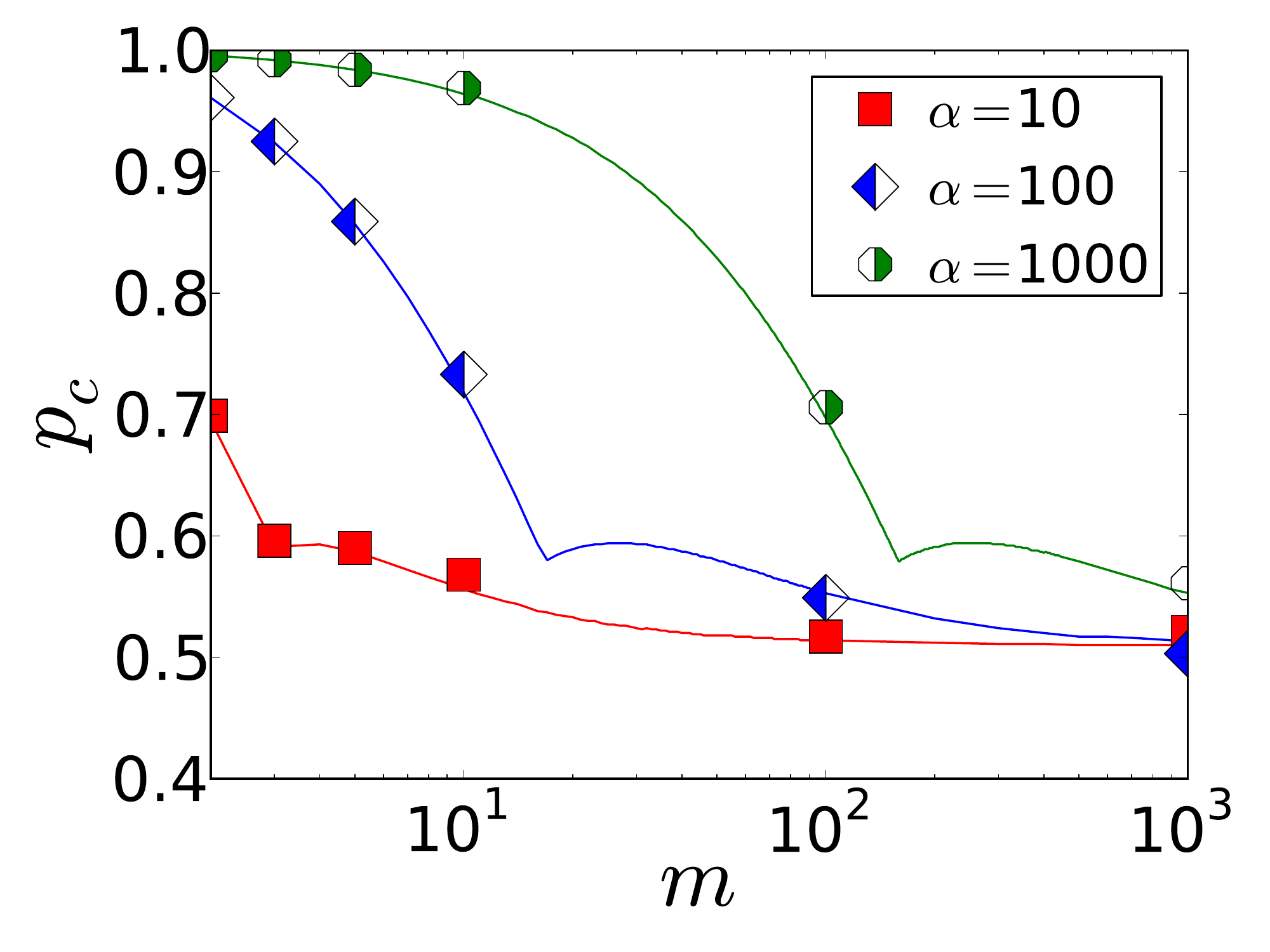} 
\caption{}
\label{fig:RR-pc}
\end{subfigure}
\hfill
\begin{subfigure}{0.45\textwidth}
\centering
\includegraphics[width=1.0\linewidth]{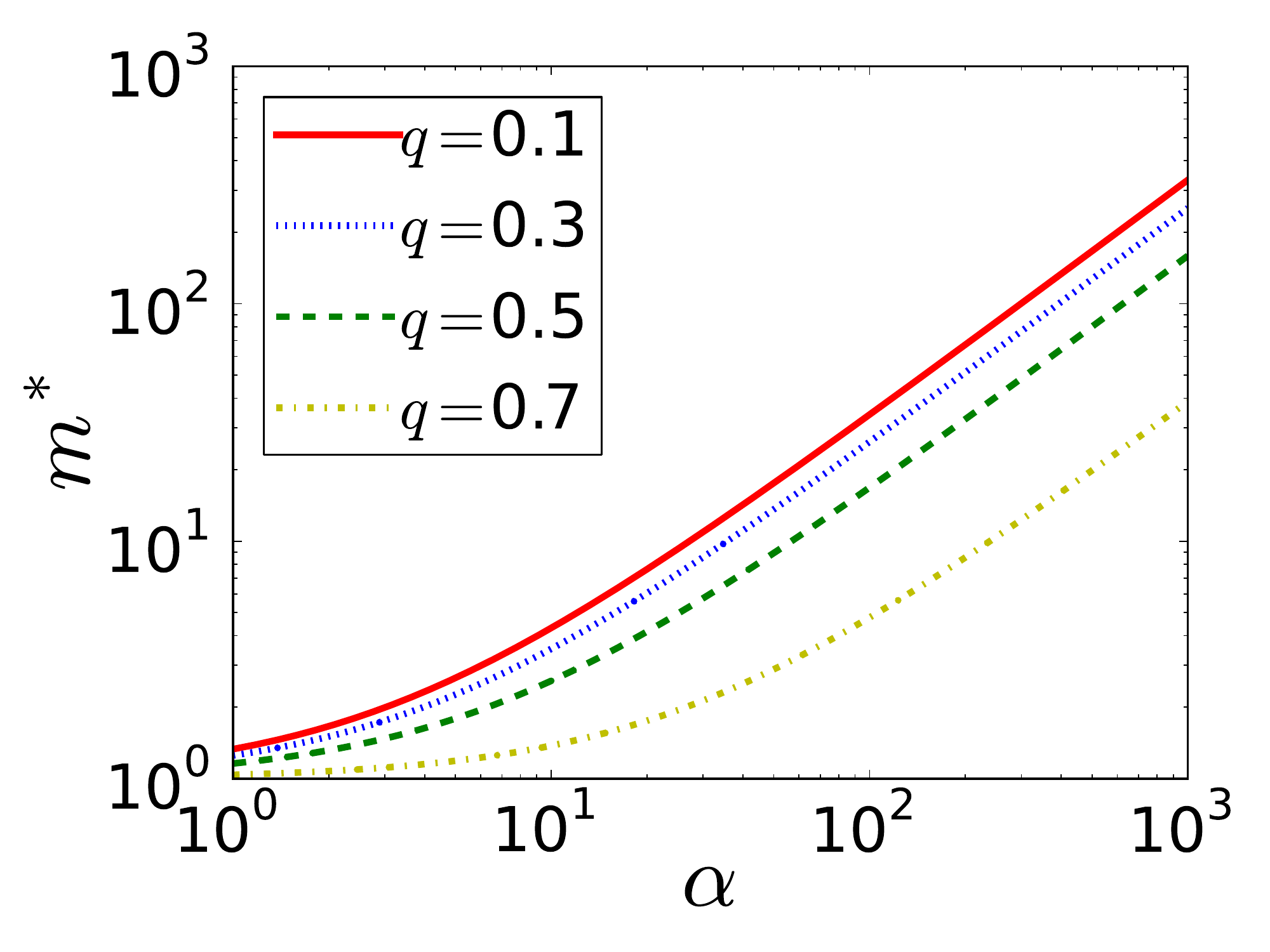} 
\caption{}
\label{fig:m*_vals-RR}
\end{subfigure}
\hfill
\caption{We show simulations and theory according to Eq. (\ref{eq:loop-pinf}) with $k=4$, $\alpha=100$ and {\bf a.} $q=0.5$ and varying $m$; and {\bf b.} $m=2$ and varying $q$. In {\bf (c)} we show $p_c$ as a function of $m$ with $q=0.5$ and several values of $\alpha$. Until the sharp kink the theory is governed by Eq. (\ref{eq:pc_r})  whereas after the kink $p_c$ changes according to Eq. (\ref{eq:pc-loop}). In {\bf (d)} we show the value of $m^*$ vs. $\alpha$ for several values of $q$ with $k=4$. }
\label{fig:RR-pinf}
\end{figure*}

The equation governing this system is once again Eq. (\ref{eq:NoN-final}) with $p_\text{dep}=(1-q+qP_\infty)$.
Substituing this into Eq. (\ref{eq:NoN-final}) gives
\begin{widetext}
\begin{equation}
P_\infty=
\begin{cases} 
\left(e^{-k_\text{inter}}(1-r)(1-e^{-k_\text{intra}P_\infty})+r(1-e^{-(k_\text{intra}+k_\text{inter})P_\infty})\right)(1-q+qP_\infty) & 0<r<1 \\
\frac{p}{m}(1-e^{-k_\text{intra}mP_\infty})(1-q+qmP_\infty) & r<0.
\end{cases}
\label{eq:loop-pinf}
\end{equation}
\end{widetext}
We compare simulations and theory according to Eq. (\ref{eq:loop-pinf}) in Fig. \ref{fig:RR-pinf}. We note that here if the system undergoes two transitions, the first one (at a higher $p$) is abrupt and the second one is either continuous or abrupt. For the parameters we used, looplike networks of networks undergo a second order transition \cite{gao-prl2011, gao-general-net}, yet for different values of $k_\text{tot}$ and $q$ they can undergo an abrupt transition. For looplike NoNs, the number of networks, $n$, does not play a role but increasing $q$ the fraction of interdependent nodes, weakens the resilience of the system (see Fig. \ref{fig:RR-varyq}). We further note that for looplike NoNs there is a maximum coupling, $q_\text{max}$ above which the entire network collapses even for $p\rightarrow 1$. Our value for $q_\text{max}$ will be the same as that of Gao et al. \cite{gao-general-net}, namely $q_\text{max}=1-1/k_\text{tot}$ since for $p=1$ our networks respond to random percolation like ER networks. Indeed, in Fig. \ref{fig:RR-varyq} we do not show higher values of $q$ since the system does not survive even for $p=1$ if $q>1-1/k_\text{tot}$. For example, with $k_\text{tot}=4$ (the value in our plots) we obtain a maximum coupling of $q=0.75$, above which the system collapses even for $p\rightarrow1$. 

Following the analysis in Chap. II we can find $p_c$ using Eq. (\ref{eq:r-c}). In this case $p_\text{dep}=(1-q+qP_\infty)$, which we will substitute into Eq. (\ref{eq:r-c}). Our system corresponding to Eq. (\ref{eq:pc-tree-mod}) becomes,
\begin{widetext}
\begin{align}
p_\text{dep}&=(1-q+qP_\infty) \\
P_\infty&=\left(e^{-k_\text{inter}}(1-r)(1-e^{-k_\text{intra}P_\infty})+r(1-e^{-(k_\text{intra}+k_\text{inter})P_\infty})\right)\\
r_c&=\frac{-b+\sqrt{b^2-4ac}}{2a} 
\label{eq:pc-loop}
\end{align}
\end{widetext}
with the values of $a$, $b$ and $c$ being the coefficients from Eq. (\ref{eq:r-c}). We plot the $p_c$ obtained based on the theory and compare with simulations in Fig. \ref{fig:RR-pc}.

To find $m^*$ we use the same method as was done in Chap. III for treelike networks of networks. Specifically we compare the $p_c$ at which the modules become separated and the $p_c$ for a network of networks with $k=k_\text{intra}$. We have from Gao et al. \cite{gao-general-net} that $p_c$ for a looplike NoN is $p_c=\frac{1}{k(1-q)}$. If we solve this and use the representation with $\alpha$ and $m$ rather than $k_\text{intra}$, we obtain
\begin{equation}
\frac{\alpha+m^*-1}{k_\text{tot}\alpha(1-q)}=e^{-\frac{k_\text{tot}(m^*-1)}{\alpha+m-1}}.
\label{eq:mstar-RR}
\end{equation}
The values of $m^*$, as a function of $\alpha$, based on Eq. (\ref{eq:mstar-RR}) can be seen in Fig. \ref{fig:m*_vals-RR}. Increasing $q$ decreases the robustness of the individual modules and thus decreases the value of $m^*$ since the system is more likely to collapse before the modules are separated.

\section{Discussion}
In summary, we have developed a framework for studying attacks on interdependent modular networks. 
Our results show that modular NoNs can behave significantly differently from random NoNs or spatially embedded NoNs in that they may undergo two separate percolation phase transitions. One transition occurs (at a higher $p$) where the modules become separated, and a second transition occurs when the individual modules collapse. For the case of a fully interdependent treelike NoN with bidirectional dependency links, both of these transitions are first order, whereas for a looplike NoN with unidirectional dependency links the first transition (higher $p$) is abrupt while the second is either abrupt or continuous depending on the parameters. These results might be relevant for many interdependent systems such as financial networks, biological networks and are particularly relevant for models of city infrastructure where most of the interdependence presumably occurs within a single city (regarded here as a community) even though there are connections to other cities. Another reason our attack is realistic for city infrastructure is because the interconnected nodes contain the interconnected links that are longer and therefore more likely to fail \cite{mcandrew2015robustness}.

We also note that the theoretical approach developed here can be used for many other types of targeted attack and for other network of network structures as well.

\section*{Acknowledgments}
LS and SH acknowledge the LINC (no. 289447 funded by the EC’s Marie-Curie ITN program (FP7-PEOPLE-2011-ITN)) and MULTIPLEX (EU-FET project 317532) projects, the Deutsche Forschungsgemeinschaft (DFG), the Israel Science Foundation, ONR and DTRA for financial support. SS thanks the James S. McDonnell Foundation 21st Century Science Initiative - Complex Systems Scholar Award (grant 220020315) for financial support.

\FloatBarrier
\bibliographystyle{naturemag}
\bibliography{paper}

\end{document}